\documentstyle[psfig]{mn}

\def\simg{\mathrel{%
      \rlap{\raise 0.511ex \hbox{$>$}}{\lower 0.511ex \hbox{$\sim$}}}}
\def\siml{\mathrel{%
      \rlap{\raise 0.511ex \hbox{$<$}}{\lower 0.511ex \hbox{$\sim$}}}}
 \def\lta{\siml}

\def\cm3{\,{\rm cm^{-3}}} \def\ee{e^\pm} \def\ie{i.e$.$ } \def\eg{e.g$.$ }
\def\etal{et al$.$ } \def\next{n} \def\npair{\tilde{n}} \def\epsilon{\varepsilon}
\def\eq{eq$.$} \def\eqs{eqs$.$} \def\reference{\bibitem}

\begin{document}

\title[Pair Wind in GRBs and Its Effect on the Early Afterglow Emission]
    {Creation of Electron--Positron Wind in Gamma-Ray Bursts and Its Effect 
       on the Early Afterglow Emission}

\author[P. Kumar and A. Panaitescu]{P. Kumar$^{1}$ and A. Panaitescu$^{2}$ \\ 
    $^{1}$Astronomy Department, University of Texas, Austin, TX 78731, 
          e-mail: pk@astro.as.utexas.edu \\
    $^{2}$Dept. of Astrophysical Sciences, Princeton University, Princeton, 
          NJ 08544, e-mail: adp@astro.princeton.edu }

\maketitle

\label{firstpage}

\begin{abstract}

 We calculate the creation of electron--positron pairs in Gamma-Ray Bursts
(GRBs) resulting from the collision between scattered and outward moving
gamma-ray photons. The number of pairs exceeds the number of ambient
medium electrons encountered by the GRB ejecta up to $\sim 10^{16}$ cm from
the center of explosion. The shock resulting from the interaction of the
ejecta with the pair-wind may brighten the afterglow synchrotron emission 
during the first few minutes. Even without this effect, the peak intensity 
of the optical afterglow increases with the density of the surrounding medium. 
Therefore, observations of the optical flux at early times constrain the density 
of the circumburst medium. If the electron and magnetic field energies behind 
the forward shock sweeping-up the pair-wind and the circumburst medium are 
as inferred from fits to the broadband afterglow emission at $0.5-100$ days, 
then the current upper limits on the optical counterpart emission, set by the 
ROTSE and LOTIS experiments, indicate that the circumburst medium within 0.01 
pc is less dense than $100\cm3$ or, if a wind, corresponds to a progenitor 
mass-loss to wind speed ratio below $10^{-6}\, M_\odot/{\rm yr}/ (10^3\, {\rm km/s})$.

\end{abstract}


\section{Introduction}

The $\gamma$-photons produced in Gamma-Ray Busts (GRBs) are scattered by electrons 
in the medium surrounding the explosion, their interaction with the outward moving 
photons creates an electron and positron pair if the photons energy in their center 
of mass frame exceeds $2\,m_e c^2$. These pairs scatter more photons, leading to an 
exponential growth of their density. The pairs are accelerated by scattering the 
gamma-ray burst photons.

The process of pair-creation in GRBs has been previously considered by
Thompson \& Madau (2000), M\'esz\'aros, Ramirez-Ruiz \& Rees (2001), and
Beloborodov (2002). The resulting pair-wind moves at a relativistic
velocity and the pairs dominate the number density of charged particles
up to a distance $10^{16}\, (E_\gamma/10^{53})^{1/2}$ cm from the center 
of explosion, where $E_\gamma$ is the isotropic-equivalent GRB energy release.
The purpose of this paper is to consider the effect of pairs created by GRB
pulse on early afterglow emission.

In \S\ref{dynamics} we consider all of previously discussed physical processes for 
electron-positron pair creation by GRB photons. We include in our calculations the 
overtaking of pair plasma by the material ejected in the explosion, which can 
substantially reduce the number of pairs produced. In \S\ref{aglow} we consider the 
effect of pairs on early afterglow emission, and calculate for various external 
densities and ejecta Lorentz factor the optical synchrotron emission produced by the 
shock that heats up the pair-wind when it collides with the GRB ejecta. The non-detection
of early optical afterglow emission brighter than magnitude $R \sim 13$ by the 
ROTSE and LOTIS experiments can be used to provide an upper limit on the density 
of the circumburst medium.

\section{Dynamics of the Electron--Positron Wind}
\label{dynamics}

The $\ee$ enrichment is calculated from the energy spectrum of the incident
and scattered photons. We approximate the GRB energy spectrum as a power-law
$F(\epsilon) = F_p (\epsilon/\epsilon_p)^{-\alpha}$ for $\epsilon > \epsilon_p$.
  For most of the 5,500 spectra of the 156 bright BATSE bursts analyzed by 
  Preece \etal (2000), the spectral slope $\alpha$ above the $\epsilon_p$ break is
  between 0.5 and 2.5, with the peak of the distribution at 1.3, while $\epsilon_p$
  is between 70 keV and 800 keV, its distribution peaking at 250 keV. Assuming a
  burst redshift $z=1$, the corrected break energy $\epsilon_p$, measured in
  $m_e c^2$ units is between 0.3 and 3, with a peak at 1.

The flux of scattered photons at some distance $x$ from the leading edge
of the gamma-ray pulse, is calculated numerically by integrating over the 
scatterings that occurred from $x=0$ to the current $x$ and over the
primary photon spectrum and deflection angle, taking into account the photon
attenuation during its propagation.

\subsection{Planar Geometry}
\label{planar}

 In planar geometry, the laboratory frame number density $n_\pm$ of the created
pairs and their streaming velocity $\beta c$ are given by
\begin{equation}
 {d\over dx} \left[ n_\pm (1-\beta) \right] = 2\,\frac{dn_\pm}{dt} \;,
\label{dn}
\end{equation}
\begin{equation}
 {d\over dx} \left\{ \left[ 2n_\pm m_e + n_p (m_e+m_p) \right]
     \gamma\beta(1-\beta)c^2\right\}=\frac{dP_{sc}}{dt} + \frac{dP_\pm}{dt} \;,
\label{dp}
\end{equation}
respectively, where $t$ is time measured in the laboratory frame, $n_p$ is the
laboratory frame number density of the swept-up external medium electrons,
$m_p$ is the proton mass, and $\gamma$ is the pair-wind Lorentz factor.
The right-hand side of equation (\ref{dn}) represents the rate of pair creation
from scattered and primary photons, while the two terms in the rhs of equation
(\ref{dp}) represent the rate of momentum deposition by pair formation and by
photon scattering, respectively. Conservation of the number of external electrons
implies that
\begin{equation}
  {d\over dx} \left[ n_p (1-\beta) \right] = 0 \;,
\label{dnp}
\end{equation}
\ie $n_p= \next/(1-\beta)$, $\next$ being the density of the unperturbed
circumburst medium.

 The most general form of the rhs's of equations (\ref{dn}) and (\ref{dp}) are
given by Beloborodov (2002) (see his equations [10] and [16]). For analytical
purposes we simplify them by making two approximations. First, we ignore
the angular dependence of the photon scattering probability and consider
that an electron (or positron) at $x$ deflects an incident photon only by the
average deflection angle $\cos^{-1} \beta(x)$. This approximation
reduces the integral of the scattered flux from one over the wind volume to
one over a surface from which scattered photons
arrive at the same time at $x$ (where $\ee$ creation takes place).
Secondly, we consider that the e$^\pm$ pairs are produced mostly by photons
near the peak of the spectrum, \ie in the integral over the spectrum of primary
photons we approximate by unity the term $1+\epsilon/\gamma$, where $\epsilon$
is the photon energy in $m_e c^2$ units.
This approximation is justified by that photons with $\epsilon > \gamma$
are scattered in the Klein-Nishina regime, thus the scattering probability is
reduced. Taking into account that the GRB photon spectrum above $m_e c^2$
is typically steeper than $\epsilon^{-2}$, the latter approximation is more
suitable at locations $x$ where the wind is relativistic ($\gamma \gg 1$).

 With the above approximations, equation (\ref{dn}) for the wind number density
becomes
\begin{eqnarray}
{d\over dx} \left[ \npair (1-\beta)\right] = {2^{-\alpha+1} \phi(\alpha)\over 
   \lambda^2} \int\limits_0^x dx'\, \npair (x')  {1-\beta(x')\over 1+ 
   \beta(x')} \times \nonumber \\
  \int\limits_{\epsilon_p}^{ \epsilon_{kn}(x')} {d\epsilon \over \epsilon}
       f(\epsilon) f(\epsilon_t) \quad, \quad
   f(\epsilon) = \left(\frac{\epsilon}{\epsilon_p}\right)^{-\alpha} \;,
 \label{dndxa}
\end{eqnarray}
where $\npair = 2\, n_\pm + n_p$ is the total lepton density, and
$\phi(\alpha) \simeq (7/12) (1+\alpha)^{-5/3}$ (Svensson 1987),
\begin{equation}
  \lambda = {c \over F_p\sigma_T} \;,
\label{lambda}
\end{equation}
$\sigma_T$ being the Thomson scattering cross-section, and
\begin{equation}
  \epsilon_t = {2(1+\beta)\over \epsilon (1-\beta)} \;,
   \quad
  \epsilon_{kn} \simeq \gamma(1+\beta) \;,
\end{equation}
are the photon threshold energy for $\ee$ formation and the Klein-Nishina
``cut-off" photon energy, respectively. Note that for $\epsilon_p < 2$
(\ie the peak of the GRB $\nu F_\nu$ spectrum is below $\sim 1$ MeV),
$\epsilon_t$ is larger than $\epsilon_p$.

 With the same approximations, equation (\ref{dp}) for the wind momentum density
becomes:
\begin{displaymath}
 {d\over dx} \left[\left(\npair+{m_p\over m_e}n_p\right) \gamma\beta(1-\beta)
          \right] = {\npair(1-\beta)\over \lambda (1+\beta)}
     \int\limits_{\epsilon_p}^{\epsilon_{kn}(x)} d\epsilon\, f(\epsilon)
\end{displaymath}
\begin{equation}
      +   {\phi(\alpha)\over \lambda^2} \int\limits_0^x dx' \; \npair(x')
           {1-\beta' \over 1+ \beta'}
           \int\limits_{\epsilon_p}^{ \epsilon_{kn}(x')} {d\epsilon \over
           \epsilon} \, f(\epsilon) f(\epsilon_t) \; p_{\pm} \;,
\label{dPdxa}
\end{equation}
where
\begin{equation}
   p_\pm = {\epsilon\;\beta' \over 1+\beta'} + {\phi(\alpha-1) \over
           \epsilon\; \phi(\alpha)} {1+\beta'\over 1-\beta'}\;,
           \; \beta' \equiv \beta(x') \;,
\end{equation}
is the total radial momentum of the scattered and primary photons that form
a $\ee$ pair. Equations (\ref{dndxa})-(\ref{dPdxa}) are similar to
equations (42)-(47) of Beloborodov (2002).

 For a power-law photon spectrum, equation (\ref{dndxa}) leads to
\begin{equation}
 {d\over dx} [\npair (1-\beta)] = {\phi(\alpha) \over 2^{\alpha-1}}
       {\epsilon_p^{2\alpha} \over \lambda^2}
    \int\limits_0^x dx' \, \frac{ \npair (x') \ln (\epsilon_{kn}/\epsilon_p) }
     { [(1+\beta') \gamma']^{2\alpha+2} } \;.
 \label{dndxb}
\end{equation}
The integrand in the above equation decreases rapidly with $\gamma$, most of
the contribution to the integral coming from $\gamma \lta 2$. Physically, this
means that pairs are created by photons scattered by mildly or non-relativistic
pairs; the $\ee$ threshold for photons scattered by relativistic $e^\pm$ is much
greater because these photons are moving within an angle $\gamma^{-1}$ relative
to the primary photons.

 As shown in Beloborodov (2002) the momentum deposition in the e$^\pm$ wind
through pair creation is at most comparable to that transferred by photon
scattering. For simplicity, we retain only the scattering term in equation
(\ref{dPdxa}), reducing it to
\begin{equation}
{d\over dx} \left[ {(\npair m_e+n_p m_p)\beta\over \gamma(1+\beta)}\right] =
  \frac{m_e \epsilon_p}{(\alpha-1)\lambda}\,\frac{\npair}{(1+\beta)^2\gamma^2}\;.
\label{dPdxb}
\end{equation}

\subsubsection{Non-Relativistic Wind}

In the $\beta \ll 1$ limit, equations (\ref{dndxb}) and (\ref{dPdxb}) lead to
an exponential growth of the pair density and velocity:
\begin{equation}
 \npair (x) = \next e^{x/x_\pm} \;,\;
 \beta(x) = \frac{\epsilon_p}{(\alpha-1)} \frac{m_e}{m_p} \frac{x_\pm}{\lambda}
             e^{x/x_\pm} \;,
\label{nb}
\end{equation}
for $x > x_\pm$, where
\begin{equation}
 x_\pm = \sqrt{ \frac{ 2^{\alpha-1}} {\phi(\alpha)}} \, \frac{\lambda}
          {\epsilon_p^\alpha} \stackrel{\alpha=1.5}{\approx}
         3\,\epsilon_p^{-3/2} \lambda \;,
\end{equation}
is the length-scale for pair loading.
For a GRB of isotropic-equivalent output energy $E_\gamma$, equation (\ref{lambda})
gives for the acceleration length-scale at radius $r$
\begin{equation}
 \frac{x_\pm}{\Delta} = \frac{\pi}{\alpha-1} \sqrt{\frac{2^{\alpha+3}}
       {\phi(\alpha)}} \frac{m_e c^2}{\sigma_T}
        \frac{r^2}{\epsilon_p^{\alpha-1} E_\gamma} \stackrel{\alpha=1.5}{\approx}
       0.1 \frac{r_{16}^2}{E_{\gamma,53} \epsilon_p^{1/2}} \;,
\label{xpm}
\end{equation}
where $\Delta = cT/(1+z)$ is the geometrical thickness of the GRB front, $T$
being the observed burst duration, $z$ the GRB's redshift, and the usual
scaling $X_n = X/10^n$ was used.

 From equation (\ref{nb}) we see that the wind acceleration length-scale 
$x_{acc}$, defined by $\gamma (x_{acc}) = 2$ is a factor of a few larger 
than $x_\pm$. A simple estimate of $x_{acc}$ can be obtained by equating
the momentum deposited by photon scattering $\npair_{acc} (\sigma_T x_{acc})
(F_p \epsilon_p m_e c^2/c^2) = \npair_{acc}\epsilon_p(m_e c)(x_{acc}/\lambda)$ 
(considering that all photons have the same energy $m_e c^2 \epsilon_p$)
with the momentum $(\npair_{acc} m_e +
n_{p,acc} m_p) c \sim \next m_p c$ of the medium (assuming that the wind mass
density is dominated by the protons, \ie $\npair (x_{acc}) \ll n_p(x_{acc})
(m_p/m_e)$, and that $n_p(x_{acc}) \approx \next$):
\begin{equation}
   {x_{acc} \over \lambda} e^{x_{acc}/x_\pm} = 
    \frac{m_p}{m_e \epsilon_p^2} \quad \rightarrow \quad
     x_{acc} \stackrel{\epsilon_p=1}{\approx} 5\, x_\pm \;.
\label{xacc}
\end{equation}
Then equation (\ref{xpm}) shows that the wind is accelerated to a
relativistic speed for radii smaller than
\begin{equation}
 r_{end} \stackrel{\alpha=1.5}{\sim} 10^{16} \epsilon_p^{1/4} E_{\gamma,53}^{1/2}\;
         {\rm cm} \;.
\label{rend}
\end{equation}

\subsubsection{Relativistic Wind}

 For $x \gg x_{acc}$ we can solve the pair density and momentum equations
together to obtain $\npair(x)$ \& $\gamma(x)$. We consider a power-law 
solution of the form
\begin{equation}
\npair(x) = A_1 \left({x \over x_{acc}} \right)^{a_1}, \quad\quad \gamma(x)
    ={A_2\over 2}\left({x\over x_{acc}}\right)^{a_2}.
\label{sola}
\end{equation}
Substituting this into equations (\ref{dndxb}) and (\ref{dPdxb}), and
assuming that the mass density is dominated by protons i.e., the number
of pairs per proton is much less than $m_p/m_e$, we obtain
\begin{equation}
(a_1 - 2a_2) A_2^{2\alpha} y^{a_1-2a_2-1} = 2\eta^2\int_{x_{min}}^x {dx'\over
  x_{acc} } \, y'^{^{a_1-2(\alpha+1)a_2}},
\label{dndxc}
\end{equation}
\begin{equation}
{a_2 n m_p y^{a_2-1}\over x_{acc}} = {m_e\epsilon_p\over \lambda (\alpha-1)}
    \left({A_1\over A_2^3}\right) y^{a_1-2 a_2},
\label{dPdxc}
\end{equation} 
where $y\equiv x/x_{acc}$, $\eta\equiv x_{acc}/x_\pm$ is a constant 
independent of $x$, and $x_{min}$ is determined by the condition that 
the pair creation optical depth for a photon scattered at $x_{min},
$ of energy $\sim \epsilon_p$, is unity between $x_{min}$ and $x$; $x_{min}$
is determined by the following equation
\begin{equation}
\tau_{\gamma\gamma}\approx (x-x_{min}) {\sigma_T f(\epsilon_t)\over m_e c^3}
 \approx 1, \quad \epsilon_t\approx {2(1+\beta)\over \epsilon_p(1-\beta)}
  \approx {8\gamma^2\over \epsilon_p}.
\end{equation}
By substituting (\ref{sola}) in the above equation we find $x_{min}$ explicitly
\begin{equation}
\tau_{\gamma\gamma}\approx {x-x_{min}\over 2^{\alpha}\lambda}
   \left({\epsilon_p\over 2\gamma}\right)^{2\alpha}\approx
   {x\, \epsilon_p^{2\alpha}\over 2^\alpha\lambda A_2^{2\alpha} }
   \left({x_{acc}\over x_{min}}\right)^{2\alpha a_2} \approx 1,
\end{equation}
or
\begin{equation}
{x_{acc}\over x_{min}}\approx {\epsilon_p^{1/2a_2} (\eta y)^{1/(2\alpha a_2)}
   \over A_2^{1/a_2} [\phi(\alpha) 2^{\alpha+1}]^{1/(4\alpha a_2)} }.
\label{xmin}
\end{equation}
It follows from equation (\ref{dPdxc}) that
 \begin{equation}
3a_2 = a_1 + 1
\label{rela}
\end{equation}
Furthermore, there are two cases to be considered for the integral in
equation (\ref{dndxc}) -- one of which is $a_1-2(\alpha+1)a_2>-1$.
The integral in this case is
dominated by the upper limit, and by equating the exponents of $y$
on the two sides of the equation we find $\alpha a^2 = 1$. This together
with equation (\ref{rela}) implies that $\alpha<1/2$, which is not a case
of interest for GRBs since the high energy spectral index for GRBs has
$\alpha>1$. The other possibility is that $a_1-2(\alpha+1)a_2<-1$, and 
in this case pair production is dominated by photons scattered at the smallest 
$x$ i.e. at $x_{min}$. Making use of the expression for $x_{min}$ (eq. 
\ref{xmin}) in equation (\ref{dndxc}), and setting the exponent of $y$ to zero
results in
\begin{equation}
a_1 - 2a_2 = {a_1 - 2a_2 + 1\over 2\alpha a_2}
\label{relb}
\end{equation}
Solving for $a_1$ and $a_2$ from equations (\ref{rela}) and (\ref{relb})
we find
\begin{equation}
 a_1 = 2 + 3/(2\alpha), \quad\; a_2 = 1 + 1/(2\alpha)
\label{solb}
\end{equation}
In other words the solution for $\npair$ and $\gamma$ are given by
\begin{equation}
  \npair(x) = A_1 \left( {x\over x_{acc}}\right)^{2 + {3\over 2\alpha}},
  \quad\quad \gamma(x) = {A_2\over 2} \left({x\over x_{acc}}\right)^{1 
   + {1\over 2\alpha} },
\label{sol}
\end{equation}
with $A_1\sim n\exp(\eta)$, the density at $x_{acc}$, and $A_2\sim 4$.
A somewhat more accurate expression for $A_1$ and $A_2$ can be obtained 
by combining equations (\ref{dndxc}), (\ref{dPdxc}) \& (\ref{solb}). The 
result in Beloborodov (2002) is a special case ($\alpha=1$) of the solution 
presented here.

\subsection{Spherical Expansion and Saturation of Wind Acceleration}
\label{spherical}

 Equations (\ref{dndxa}) and (\ref{dPdxa}) were derived under the assumptions
of planar geometry and stationary solution in the $x$-coordinate. Due to the
spherical expansion, the rate of momentum deposition decreases as $r^{-2}$,
while the rate of pair creation decreases as $r^{-4}$. The effect of spherical
expansion is easily accounted for in a numerical treatment by decreasing the
photon flux: $F(r) = F(r_\gamma)(r/r_\gamma)^2$, where $r_\gamma$ is the
radius at which the GRB photons were emitted, and by adjusting similarly at
each timestep the laboratory frame density of the created pairs $n_\pm (x)$
and accelerated external electrons $n_p(x)$.

 The Lorentz factor $\gamma_{sph}$ and pair density $\npair_{sph}$ at which the
pair creation process saturates due to spherical expansion 
({\it photon dilution})
can be calculated approximately as the value of these variables at the 
coordinate $x_{sph}$ corresponding to a twofold increase
of radius $r$ at which the external medium entered the GRB front: $r =
\int_0^{x_{sph}} dx' (1-\beta')^{-1}$. Using equation (\ref{sol}), one obtains
\begin{equation}
 x_{sph} \simeq \left[ {(3+\alpha^{-1})r\over 8 x_{acc} }
    \right]^{\alpha/(1+3\alpha)} x_{acc}
\label{xsph}
\end{equation}
and
\begin{equation}
 \npair_{sph}  \simeq A_1 \left[ {(3+\alpha^{-1})r\over 8 x_{acc} }
    \right]^{ {4\alpha + 3\over 2(1+3\alpha)} }
\label{ndil}
\end{equation}
\begin{equation}
 \gamma_{sph} \simeq {A_2\over 2} \left[ {(3+\alpha^{-1})r\over 8 x_{acc} }
    \right]^{ {1 + 2\alpha\over 2(1+3\alpha)} }
\label{gdil}
\end{equation}
where
\begin{equation}
 x_{acc} \simeq {1.5\times10^{-3} r_{15}^2 \eta \over E_{\gamma,53} 
    \epsilon_p^\alpha \sqrt{\phi(\alpha) 2^{1-\alpha}} }
\label{xacca}
\end{equation}
At $x > x_{sph}$, the wind density decreases as $\npair \propto R^{-1/2} \propto
x^{-2}$, where $R \sim 2\, \gamma_{sph}^2 x$ is the front radius corresponding
to coordinate $x$. 

 The Lorentz factor of pair-wind is also limited by the non-zero transverse
momentum of gamma-ray photons: the angle $\theta$ of a GRB photon w.r.t. the
radial direction is $\theta= \Gamma^{-1} (r_\gamma/r)$, where $\Gamma$ 
is the Lorentz
factor of the relativistic ejecta which produced the GRB photons at $r_\gamma$.
Consequently, when the pair-wind reaches the Lorentz factor $\gamma_\theta =
\Gamma (r/r_\gamma)$ the photon field becomes isotropic in the frame of the
moving scatterer and there is no further net transfer of momentum from photons
to the wind. As shown by Beloborodov (2002), this effect yields a multiplying
factor $[1-(\gamma/\gamma_s)^4]$ in the $P_{sc}$ term in equation (\ref{dp}),
where $\gamma_s = \Gamma (r/r_\gamma)$. From equation (\ref{sol}), the 
acceleration saturates at coordinate $x_\theta/x_{acc} \simeq
(2\Gamma r/A_2 r_\gamma)^{2\alpha/(1+2\alpha)}$, where the wind 
density is $\npair_\theta \simeq A_1 (2\Gamma r/A_2 r_\gamma)^{(3+4\alpha)
  /(1+2\alpha)}$.

 Using the above equations for $r(x)$, $x_{sph}$, and $x_\theta$, it can 
be shown that the angular dispersion of GRB photons is more important than
the {\it photon dilution} effect (\ie $x_\theta < x_{sph}$) at
radius $r$ less than
\begin{equation}
 r_* \simeq \left( {A_2 r_\gamma\over 2\Gamma}\right)^{ {2(1 + 3\alpha)\over
   3+8\alpha} } \left[ {2.8\times10^{21} E_{\gamma,53} (3+\alpha^{-1})
     \over T(1+z)^{-1}\eta \epsilon_p^{-\alpha} \{
   \phi(\alpha) 2^{1-\alpha}\}^{-1/2} }
   \right]^{ {1 + 2\alpha\over 3+8\alpha} }
\end{equation}
Thus, if the GRB emission is produced in internal shocks occurring at $r_\gamma
\sim 2\, \Gamma^2 c \delta t \sim 6\times 10^{13}\, \Gamma_2^2 \delta t_{-1}$ cm, 
where $\delta t \sim 0.1$ s is the GRB variability timescale, the lepton density 
and Lorentz factor of the medium in the wake of the GRB front, at $r\lta10^{14}$cm 
and $\alpha=1.5$, are
\begin{equation}
 \npair_\theta \sim 3 \times 10^7\, \next \left( \frac{r_{14}}{\Gamma_2 \delta t_{-1}}
          \right)^3 \;, \;
 \gamma_\theta \sim 170\, \frac{r_{14}}{\Gamma_2 \delta t_{-1}} \;.
\label{ngcol}
\end{equation}
As shown in equation (\ref{dnp}), the acceleration of the external electrons
compresses them to a density $n_{p,\theta} = 2\,\gamma_\theta^2 \next$,
therefore the pair-enrichment factor $\mu \equiv \npair/n_p$ (\ie the number
of $e^-$ and $e^+$ created for each external electron) is $\mu_\theta \sim
500\; r_{14}/(\Gamma_2 \delta t_{-1})$.

 The medium encountered at $r > r_*$ is pair-loaded and accelerated up to
the density and Lorentz factor given by equations (\ref{ndil}) and (\ref{gdil}).
By substituting $x_{acc}(r)$ from equations (\ref{xacc}) and (\ref{xpm}),
we obtain
\begin{equation}
 \npair_{sph} \simeq A_1 \left[ {2.8\times10^6 (3+\alpha^{-1}) E_{\gamma,53}
   \sqrt{\phi(\alpha) 2^{1-\alpha}}\over T(1+z)^{-1} r_{15}\eta
    \epsilon_p^{-\alpha} }\right]^{ {3 + 4\alpha\over 2+6\alpha} }
\label{nsph}
\end{equation}
\begin{equation}
 \gamma_{sph} \simeq {A_2\over2} \left[ {2.8\times10^6 (3+\alpha^{-1}) 
  E_{\gamma,53}\sqrt{\phi(\alpha) 2^{1-\alpha}}\over T(1+z)^{-1} r_{15}\eta
    \epsilon_p^{-\alpha} }\right]^{ {1 + 2\alpha\over 2+6\alpha} }.
\label{gsph}
\end{equation}
Note that $A_1=\npair_{sph}(x=x_{acc})\simeq n \exp(\eta)$, and $A_2\sim 4$.
The pair-enrichment factor $\mu_{sph}=\npair_{sph}/(2n\gamma_{sph}^2)\propto 
(E_{\gamma,53}\epsilon_p^\alpha/T r_{15})^{1/(2+6\alpha)}$.
If $x_{sph} > \Delta$ the trailing edge of the GRB front is encountered
before the wind acceleration and pair loading is saturated.
This happens for electrons entering the GRB front at radii larger than
\begin{equation}
 r_{**} \sim {10^{15} (6.6\times10^2)^{ {1 + 3\alpha\over 2+5\alpha} }
    \left[T/(1+z)\right]^{{\alpha\over 2+5\alpha}} \over
    \left[2.8\times10^6(3+\alpha^{-1})\right]^{{\alpha\over 
    2+5\alpha}} } \left[ { E_{\gamma,53} \epsilon_p^{\alpha}\over
    \eta \phi(\alpha)^{-1/2} }
    \right]^{ {1 + 2\alpha\over 2+5\alpha} }
   {\rm cm}.
\end{equation}
In this case, substituting $x=\Delta$ in equation (\ref{sol}), yields
\begin{equation}
 \npair_\Delta \simeq n\exp(\eta) \left[ {6.6\times10^2 E_{\gamma,53}
    \sqrt{\phi(\alpha) 2^{1-\alpha}}\over r_{15}^2\eta\epsilon_p^{-\alpha} }
    \right]^{ {3 + 4\alpha\over 2\alpha} }
\label{nd}
\end{equation}
and
\begin{equation}
 \gamma_\Delta \simeq {A_2\over 2} \left[ {6.6\times10^2 E_{\gamma,53}
    \sqrt{\phi(\alpha) 2^{1-\alpha}}\over r_{15}^2\eta\epsilon_p^{-\alpha} }
    \right]^{ {1 + 2\alpha\over 2\alpha} }
\label{gd}
\end{equation}
for the maximal wind Lorentz factor and pair density, thus the pair-enrichment
factor is $\mu_\Delta = \npair_\Delta/(2 n\gamma_\Delta^2)\propto 
(\epsilon_p^\alpha E_{\gamma,53}/r_{15}^2)^{1/2\alpha}$.

The main results of this section is that the pair creation and acceleration
up to $r_{**} \sim 10^{15.3}$ cm is limited by the decrease in photon density
due to spherical expansion. The pair creation for 
$r_{**}<r<r_{end}\sim 10^{16}$cm continues throughout the
gamma-ray pulse and the results in this regime are the same for spherical 
and plane parallel geometries. The medium beyond $r_{end}$ is not affected
by the passage of the GRB front.

 Most of the kinetic energy of the pair-wind is acquired at $r \lta r_{rel}$;
$r_{rel}\sim 2.6\times10^{16} (E_{\gamma,53}\sqrt{\phi(\alpha) 2^{1-\alpha}}
\epsilon_p^\alpha/\eta)^{1/2}$cm, is the radius where the pair-wind Lorentz 
factor drop below 2.  Since the pair-enrichment 
factor $\mu$ does not exceed $m_p/m_e$, the wind kinetic energy is approximately
that of the entrained circumburst medium protons (or ions). Using equation
(\ref{gd}), the wind kinetic energy is found to be
\begin{equation}
 E_w = 4\pi\, \next m_p c^2 \int\limits_0^{r_{end}} r^2 \gamma_\Delta dr
   \lta  10^{47} n\,\epsilon_p^{3\alpha/2} E_{\gamma,53}^{3/2} \; {\rm erg} \;,
\label{Ewind}
\end{equation}
and therefore the fraction of energy in gamma-ray pulse converted to the 
creation of a pair-wind is $10^{-6} n_0$. The overall wind Lorentz factor 
$\gamma_w = E_w/ (m_w c^2) \sim 3$, where $m_w \sim (4\pi/3)\,r_{end}^3 \next m_p$ 
is the wind mass.

\subsection{Numerical Results}
\label{numerical}

 There are two factors associated with the wind spherical expansion which have
been ignored in the analytical calculations presented in section \S\ref{dynamics}: 
the spherical dilution of the incident GRB photon flux and the interactions within 
the pair-wind. The former has been considered in \S\ref{spherical}; here we note
that another way in which it leads to a non-stationary $\npair (x)$ and $\gamma (x)$ 
is that the GRB front radius may increase significantly while a shell of external
matter, undergoing acceleration and pair-enrichment, is within the front.
Ignoring this effect is appropriate when the upper limit to the radius increase
during the front-crossing, $2 \gamma_\Delta^2 \Delta$, is less than the front
radius, $r$, which, after using equation (\ref{gd}), leads to $r > 2\times 10^{15}
T_1^{0.16}$ cm for $\alpha = 1.5$, $E_\gamma = 10^{53}$ ergs, $\varepsilon_p =1 $.
Collisions within the pair-wind occur because
the external medium entering the GRB front at smaller radius is accelerated to a 
higher Lorentz factor and may overtake a slower shell of external matter entering 
the GRB front at a larger radius. A third factor is the presence of the ejecta within 
the GRB pulse, which sweep-up the pair-wind and may drastically reduce the acceleration
through photon-scatterings if the ejecta Lorentz factor is significantly larger than 
that of the wind. Thus this factor is of importance only when the lag $\delta$ between 
ejecta and the leading edge of the GRB front, $r/(2\Gamma^2)$, is less than the pulse 
width, $\Delta$, \ie at radii $r \siml 6 \times 10^{15} \Gamma_2^2 T_1$ cm, and only 
if $\Gamma \gg \gamma (\delta)$.

 The effects discussed above are taken into account in the results shown in Figure 1. 
The external medium is discretized in shells of thickness much smaller than their 
radius, which entering the GRB front at $x=0$ and are accelerated, compressed, and 
pair-loaded as described in \S\ref{dynamics}. To include the spherical expansion, 
the GRB flux and the shell density are adjusted after each time-step. The progressive 
merging of shells (interactions within the wind) is accounted for by replacing a pair 
of colliding shells with one whose Lorentz factor and density are calculated from 
energy conservation and the jump conditions at shocks, respectively. The location of 
the ejecta is tracked, allowing for the accumulation of shocked wind shells and 
compression of the shocked ejecta.

\begin{figure}
\centerline{\psfig{figure=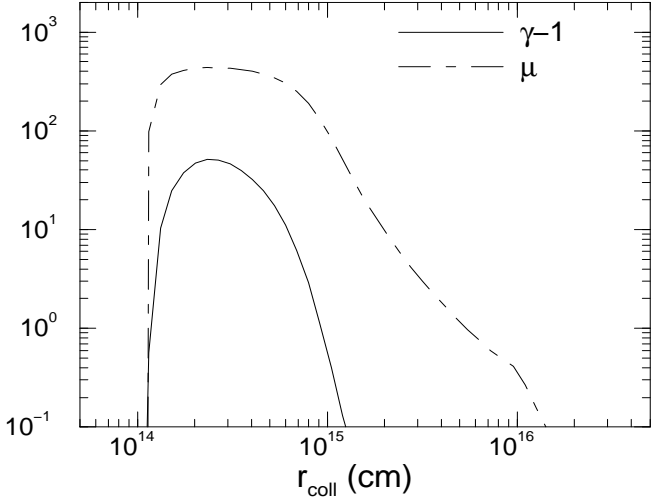,width=8.5cm,height=7cm}}
\caption{Wind acceleration and pair-loading for a GRB pulse of isotropic-equivalent
 energy $E_\gamma=10^{53}$ erg, intrinsic duration $T=10$ s and peak of $\nu F_\nu$ 
 spectrum $\epsilon_p = m_e c^2$, post-peak slope $\alpha = 1.5$ ($F_\nu \propto 
 \nu^{-\alpha}$). The GRB emission is assumed to arise at $r_\gamma \siml 10^{14}$ 
 cm from ejecta moving at $\Gamma = 100$. Calculations take into account the interactions 
 within the wind, between the wind and the GRB ejecta, and the effect of spherical
 expansion on the incident flux and wind density. The abscissa gives the radial coordinate 
 at which the ejecta--wind shell interactions take place.}
\end{figure}

 The termination of the wind acceleration while it is still within the GRB front,
due to the collision with the ejecta, is illustrated in Figure 1 by the rising
part of the wind density ($n\pm$) and Lorentz factor ($\gamma_\pm$) at $r \sim 10^{14}$
cm. As the ejecta lag more behind the front's leading edge, the external medium is
more accelerated and pair-loaded. The interactions within the wind and the decrease of
the incident flux with radius lead to a flattening of the wind Lorentz factor at 
$r = 2-4 \times 10^{14}$ cm, followed by a decrease. The ejecta drive the forward shock
into a relativistic wind up to $r_{rel} \siml 10^{15}$ cm, thus the dissipation efficiency 
of this shock is reduced by the existence of the pair-wind up to an observer time
$t_{rel} \sim (1+z) (r_{rel}/c\Gamma^2) \siml 10 \Gamma_2^{-2}$ s. 
The pair production becomes negligible at $r_{pair} \sim 10^{16}$ cm, corresponding to an
observer time $t_{pair} \sim 10 t_{rel} \siml 100 \Gamma_2^{-2}$ s. As shown in Figure
1, the pair enrichment factor decreases roughly as $\mu \propto r_{coll}^{-3}$ for
$10^{15} < r < 10^{16}$ cm. This means that most of the radiating leptons at $r = r_{pair}$ 
are the pairs formed at $r < r_{pair}$ and not the electrons originally existing in the 
circumburst medium. Then we expect the lepton enrichment to have an effect on the afterglow 
emission until later than $t_{pair}$.

\section{Early Afterglow Emission}
\label{aglow}

 A shell of circumburst medium that enters the GRB front at radius $r_{in}$
leaves the front at radius $r_{out} \sim r_{in} + \Delta/(1-\beta_\pm)$,
where $\beta_\pm$ is the shell velocity reached when it exits the GRB front,
the corresponding Lorentz factor $\gamma_\pm \equiv \gamma (x=\Delta)$ being
given by equations (\ref{ngcol}), (\ref{gsph}) or (\ref{gd}).
The GRB ejecta collide with the pair-wind at $r_c \sim [1+(\gamma_\pm/\Gamma)^2]
r_{out}$ (assuming that $\gamma_\pm \ll \Gamma$) if the GRB ejecta are outside
the front, \ie $r_c > 2\, \Gamma^2 \Delta$. If the GRB ejecta interact with the 
pair-wind while the layer is still within the GRB front, its pair-loading and 
Lorentz factor are lower. 

 In the frame of the unshocked pair-wind, the GRB ejecta move at the Lorentz
factor $\gamma_r = \Gamma \gamma_\pm (1-\beta_\pm \beta_e)$, where $\beta_e$ is
the ejecta velocity in the laboratory frame. Denoting by $\epsilon_e$ the fractional
energy of the shocked wind in $e^-$ and $e^+$, the typical lepton random Lorentz
factor is:
\begin{equation}
 \gamma_p = \frac{p-2}{p-1} \epsilon_e (\gamma_r - 1)
            \left( 1 + \frac{n_p m_p}{\npair m_e} \right)\;,
\label{gp}
\end{equation}
where $p$ is the exponent of the power-law energy distribution $dn/\gamma_e \propto
\gamma_e^{-p}$ at $\gamma_e > \gamma_p$ acquired by the leptons.

 At $r_c$ the wind density is smaller by a factor $(r_c/r_{out})^2$ than that
at $r_{out}$. The shock compression at $r_c$ increases the comoving density
of the wind by a factor $4\,\gamma_r + 3$. If the magnetic field stores a
fraction $\epsilon_B$ of the shocked wind internal energy, then the magnetic
field is:
\begin{equation}
  B^2=8\pi\,\epsilon_B (4\gamma_r+3)(\gamma_r-1)(\npair_\pm m_e + n_{p,\pm} m_p)
                (r_{out}/r_c)^2 \;,
\label{B}
\end{equation}
where $\npair_\pm \equiv \npair (x=\Delta)$ and $n_{p,\pm} \equiv n_p (x=\Delta)$.

 Figure 2 shows the evolution of the characteristic synchrotron frequency $\nu_p =
(3e/16 m_e c) \gamma_p^2 B \Gamma$ corresponding to a typical electron energy in the 
shocked wind (\eq [\ref{gp}]). As discussed in \S\ref{numerical}, the pair-wind affects 
the afterglow emission by reducing the energy per lepton in two ways: until $t_{rel} 
\siml 10\, \Gamma_2^{-2}$ s, it lowers the dissipated energy and, until some time later
than $t_{pair} \siml 100\, \Gamma_2^{-2}$ s, it increases the number of particles among 
which this energy is shared. Both these effects lower the typical lepton Lorentz factor 
$\gamma_p$, leading to a softening of the afterglow emission. As the GRB front radius 
increases, the wind is less accelerated and pair-enriched and the afterglow spectrum 
hardens, its peak frequency $\nu_p$ asymptotically reaching a constant value. Note in
Figure 2 that the softening of the spectrum depends on $\Gamma$, for higher initial
Lorentz factor, the gap between the leading fronts of the GRB pulse and ejecta is 
smaller, thus the pair-wind is less accelerated and enriched before it is swept-up. 

\begin{figure}
\centerline{\psfig{figure=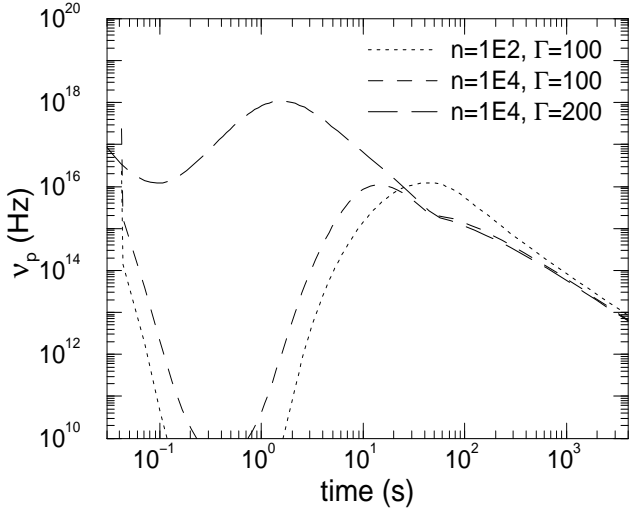,width=8.5cm,height=7cm}}
\caption{ Evolution of the peak frequency of the synchrotron spectrum $\nu_p$ 
 for three sets of parameters $(n,\Gamma)$ (external particle density and fireball 
 initial Lorentz factor). The burst is placed at redshift $z=1$ (redshift), other
 parameters being as for Figure 1. The kinetic energy of the ejecta $E_k$ was taken 
 equal to $E_\gamma$ (\ie a 50\% efficiency of the GRB emission was assumed).
 The power-law energy distribution of the shock-accelerated leptons,
 $dn/d\gamma_e \propto \gamma_e^{-p}$, has a total energy equal to a fraction 
 $\epsilon_e = 0.1$ of the post-shock energy and an index $p=2.5$. The magnetic field 
 energy is a fraction $\epsilon_B = 10^{-4}$ of the internal energy. The photon arrival 
 time is calculated for the photons moving along the line of sight toward the center 
 of explosion. Note that the pair-wind softens the afterglow spectrum up to 10 seconds.}
\end{figure}

 When the mass of the swept-up circumburst medium reaches a fraction $\sim \Gamma^{-1}$ 
of the ejecta mass, the fireball begins to decelerate. This happens at
\begin{equation}
  r_{dec} \equiv \left( \frac{3 E_k}{4\pi \next m_p c^2 \Gamma^2}\right)^{1/3}
   \sim 10^{17} \left( \frac{E_{k,53}}{\next_0 \Gamma_2^2}\right)^{1/3} \,{\rm cm} \;,
\label{rdec}
\end{equation}
where $E_k$ is the kinetic energy of the ejecta after the GRB phase, corresponding to 
an observer time $t_{dec} = (1+z) r_d /(\Gamma^2 c) \sim 700\, (E_{k,53}/\next_0)^{1/3} 
\Gamma_2^{-8/3}$ s. At $t > t_{dec}$, the fireball Lorentz factor is $\Gamma = [E_k/m(r)]^{1/2}$ 
(for adiabatic dynamics), where $m(r)$ is the mass of the energized circumburst medium. 
Due to the decreasing $\Gamma$, the afterglow spectrum softens again, the peak frequency 
evolving as
\begin{equation}
 \nu_p (t) \sim  3\times 10^{15} E_{k,53}^{1/2} \epsilon_{e,-1}^2 
                  \epsilon_{B,-4}^{1/2} t_2^{-3/2} \, {\rm Hz} \;.
\label{nup}
\end{equation}

 To obtain the afterglow optical light-curves, we integrate the synchrotron emission 
function over the electron distribution, taking into account effect of the spherical 
curvature of the wind surface on the photon arrival-time, photon frequency Doppler boost, 
and relativistic beaming. Equations (\ref{gp}) and (\ref{B}) establish the initial 
distribution of the shock-accelerated pairs and the magnetic field, for the wind dynamics 
described in \S\ref{dynamics}. We track numerically the evolution of the electron distribution 
(subject to radiative and adiabatic losses) in each shell of external matter, after it is 
accelerated and pair-loaded by the GRB front and compressed and heated by the collision with 
GRB ejecta. The radiative and adiabatic losses are also included in the calculation of the 
afterglow dynamics.

 Figures 3 and 4 illustrate the effect of the pair-wind on the early optical afterglow
for a few values of the circumburst medium density $n$ and fireball initial Lorentz factor 
$\Gamma$. To understand better this effect, we outline first the properties of the
optical light-curves expected when the radiative cooling of the $\gamma_p$-electrons and 
the pair-wind are ignored. Then the light-curve would peak at time $t_p$ when $\nu_p$ 
passes through the optical band. Equation (\ref{nup}) gives that 
\begin{equation}
 t_p \sim 400 E_{k,53}^{1/3} \epsilon_{e,-1}^{4/3} \epsilon_{B,-4}^{1/3} \, s \;
\label{tp}
\end{equation}
independent of $n$ or $\Gamma$. The optical flux at $t_p$ is the peak flux of the
synchrotron spectrum, the corresponding optical magnitude being
\begin{equation}
 R (t_p) = 14 - 2.5 \log (E_{k,53} n_2^{1/2} \epsilon_{B,-4}^{1/2}) \;,
\label{Rp}
\end{equation} 
for redshift $z=1$. Note that $R(t_p)$ increases with the external density. The 
light-curves shown in Figures 3 and 4 with thin lines are consistent with the results 
in equations (\ref{tp}) and (\ref{Rp}) for the lowest densities and Lorentz factors 
considered here. As the magnetic field $B$ and typical electron Lorentz factor increase
with $n$ and $\Gamma$ (\eqs [\ref{B}] and [\ref{gp}]), the $\gamma_p$-electrons cool 
faster for larger $n$ and $\Gamma$, losing their energy on a timescale shorter than the 
dynamical timescale ($\sim t_{dec}$). As a result, the electron distribution develops 
a tail at energies lower than $\gamma_p$, the peak of the synchrotron spectrum moves 
at a frequency below $\nu_p$, and the optical light-curve peaks at an earlier time.
The peak magnitude remains that given in equation (\ref{Rp}).

\begin{figure}
\centerline{\psfig{figure=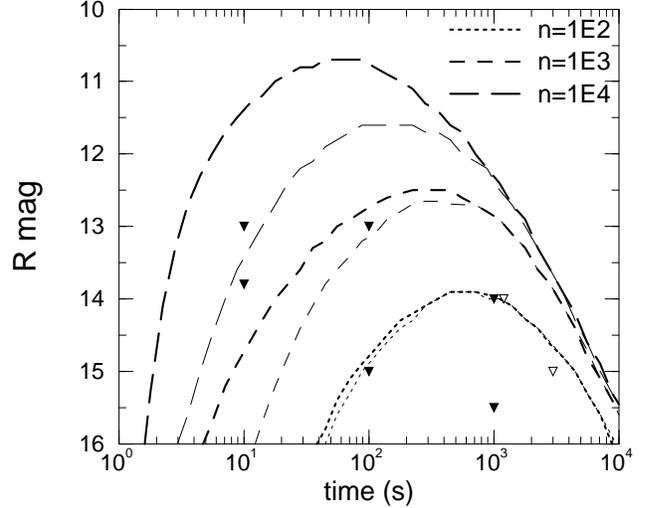,width=8.5cm,height=7cm}}
\caption{{\sl Thick lines}: afterglow optical light-curves produced for various 
 particle densities $n$ of the circumburst medium, and for a fireball initial 
 Lorentz factor $\Gamma = 100$. Other parameters are as for Figures 1 and 2. 
 {\sl Thin lines}: light-curves that would be obtained if the pair-wind were ignored.
 Note that the brightening of the afterglow emission during the first few minutes
 by the pair-wind increases with the circumburst density. Filled triangles represent
 upper limits obtained by ROTSE, open triangles are for LOTIS limits.}
\end{figure}

\begin{figure}
\centerline{\psfig{figure=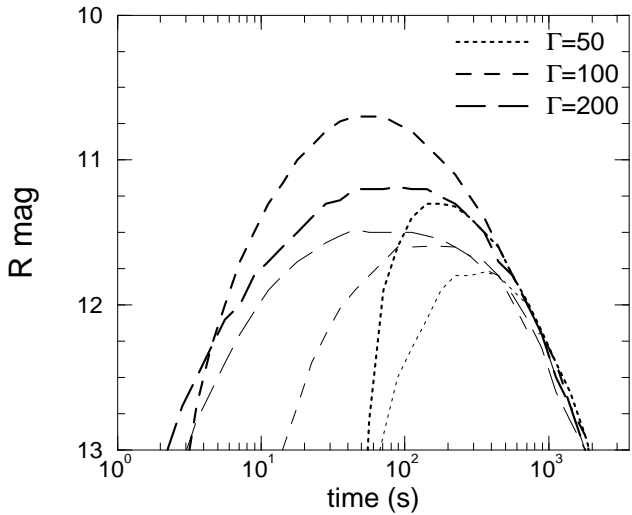,width=8.5cm,height=7cm}}
\caption{Same as in Figure 3, but for various ejecta Lorentz factors $\Gamma$
 and a fixed particle density of external medium $n=10^4 \cm3$.
 Note that for $\Gamma = 100$ the brightening produced by the pair-wind is maximal.
 For $\Gamma \simg 300$ the wind reaches only a small fraction of the maximal
 Lorentz factor before being swept-up by the forward shock driven by the GRB ejecta,
 the effect of the wind on the afterglow light-curve becoming negligible.}
\end{figure}

 If the early afterglow spectrum peaks above the optical domain (as is expected for
values of the microphysical parameters not far from those used in \eq [\ref{nup}]),
then the spectrum softening and increase of the number of radiating particles produced
by the pair wind should brighten the afterglow optical emission until a time of the
order of $t_{pair}$. Figures 3 and 4 confirm this
expectation and also show that the magnitude of the brightening induced by the pair-wind 
depends on the external density and ejecta initial Lorentz factor. As discussed above,
a denser medium and a higher Lorentz factor lead to a faster lepton radiative cooling,
which makes the optical light-curve peak at an earlier time. For the afterglow parameters 
considered in Figures 3 and 4, external densities larger than about $n \sim 10^3\,\cm3$ 
lead to light-curve peak times earlier than $t_{pair}$. In this case, the pair-wind 
brightens not only the rising part of the afterglow light-curve but also its peak, the 
effect increasing with $n$. Increasing the ejecta Lorentz factor has a similar effect;
however, for a larger $\Gamma$, the premature termination of the wind loading reduces
the brightening produced by the pair-wind.

 Summarizing the results shown in Figures 3 and 4, we can say that, without the pair-wind 
and for a set of representative afterglow parameters, the optical afterglow peak should be
brighter than $R = 13$ for $n \simg 10^3\, \cm3$, should increase by one magnitude for 
an ten-fold increase in the external density, but this peak should occur earlier than
100 seconds if $n \simg 10^4\,\cm3$. Further, for $\Gamma \siml 200$, the pair-wind 
brightens the optical afterglow peak of the optical afterglow by half of a magnitude for 
$n = 10^3\, \cm3$ and by 1 magnitude for $n = 10^4\, \cm3$; however, for $\Gamma \simg 200$,
the effect of the pair-wind is reduced.

 For the canonical afterglows parameters given in Figure 2 and $n \sim 10^4\,\cm3$, 
the optical afterglow shown in Figures 3 is brighter than the upper limit $R \simg 13$ 
at $t = 10$ s set by the ROTSE experiment (Akerlof \etal 2000, Kehoe \etal 2001) for 
the prompt optical emission of two GRBs; however, a lower density or a higher Lorentz
factor (suppressing the pair-wind brightening), would explain the non-detections. 
Similarly, the upper limits ranging from $R = 13$ to $R = 15$, obtained by ROTSE at 
$t = 100$ s for three bursts, require that $n \siml 10^3\,\cm3$ and $n \siml 10^2\,\cm3$, 
respectively, the pair-wind having a small effect at such times and for such densities. 
Finally, $n < 10^2\,\cm3$ is required by the upper limits ranging from $R = 14$ to $R=16$ 
set at $t = 10^3$ s by ROTSE for five bursts (see also Rykoff \etal 2002) and by 
$R (t > 1200s) > 14$ and $R (t < 3000s) > 15$ obtained by LOTIS for GRBs 971227 and 010921, 
respectively (Williams \etal 1999, Park \etal 2002).

 The above comparison between expectations and observations for the early afterglow emission
suggest that the circumburst density does not exceed $n = 10^2-10^3 \, \cm3$, unless
the microphysical parameters $\epsilon_B$ and $\epsilon_e$ and the kinetic energy $E_k$
have lower values than considered for Figures 3 and 4. The effect of these parameters
is illustrated in Figure 5 for $n=10^3\, \cm3$ and $\Gamma = 100$. A kinetic energy $E_k$
lower than $10^{52}$ ergs or magnetic field parameter $\varepsilon_B$ below $10^{-6}$ yield 
an early afterglow emission too dim to have been detected with the past capabilities of ROTSE 
or LOTIS, but within their reach in the future. For a lower electron energy, past detections 
were more difficult, but not impossible. We note that such low values for $E_k$ and the
$\gamma$-ray output energy $E_\gamma$ obtained by Bloom, Frail \& Sari (2001) for bursts
with known redshifts, ranging from $10^{51.5}$ to $10^{54.5}$ erg, would imply high burst
efficiencies. Furthermore, the above low values for $\varepsilon_e$ and $\varepsilon_B$ 
are below the values $\epsilon_e > 0.1$ and $\epsilon_B > 10^{-4}$ determined by Panaitescu 
\& Kumar (2002) from fits to the 0.5--100 days broadband afterglow emission of several GRB 
afterglows. 

\begin{figure}
\centerline{\psfig{figure=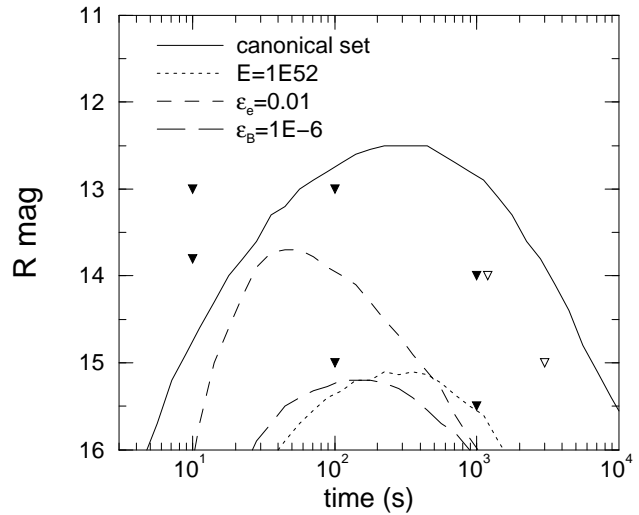,width=8.5cm,height=7cm}}
\caption{ Effect of the burst energy, electron, and magnetic field parameter on the 
 afterglow optical emission. The canonical set of parameters is: $n=10^3\,\cm3$, 
 $\Gamma = 100$, $E_k = E_\gamma = 10^{53}$ erg, $z=1$, $T=20$ s, $\epsilon_p = 
 m_e c^2$, $\alpha = 1.5$, $\epsilon_e = 0.1$, $\epsilon_B = 10^{-4}$, $p = 2.5$. 
 Varying the burst duration, the peak energy of the GRB spectrum, or its slope 
 has little effect on the resulting afterglow light-curve.}
\end{figure}

 Therefore, barring a substantial increase of the $\epsilon_e$ and $\epsilon_B$ parameters 
from $10^2-10^3$ s to $10^5$ s after the burst, the non-detections by LOTIS and ROTSE of 
GRB optical counterparts brighter than $R = 13$--16 at $t < 3\times 10^3$ s imply that 
the medium surrounding the GRB source is less dense than $n = 100\, \cm3$ up to a distance 
$r \sim 0.01$ pc. 

 Figures 6 and 7 present the afterglow light-curves expected when the circumburst medium 
is a stellar wind of density profile $n = A r^{-2}$ ejected by the GRB progenitor, as 
expected in the hypernova scenario. In contrast with the homogeneous medium case considered 
in Figures 3 and 4, the brightening produced by the pair wind is stronger (2--3 mags), occurs 
earlier, and has a weak dependence on the normalization constant $A$ (\ie the wind density), 
but a stronger dependence on the ejecta Lorentz factor. For the canonical burst parameters 
used here, optical afterglows brighter than some of the ROTSE and LOTIS limits are obtained 
even for tenuous stellar winds corresponding to a mass-loss rate to wind speed ratio of 
$10^{-6}\, M_\odot/{\rm yr}/ (10^3\, {\rm km/s})$.

\begin{figure}
\centerline{\psfig{figure=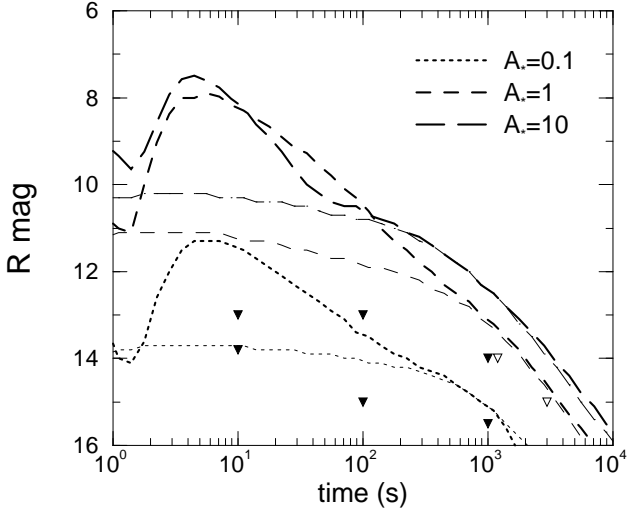,width=8.5cm,height=7cm}}
\caption{ Afterglow light-curves for the same parameters as in Figures 1 and 2, but
 for an external medium of density $n = A r^{-2}$, as expected if the GRB progenitor
 is a massive star. The legend gives the constant $A$ divided by that corresponding 
 to a mass-loss rate of $10^{-5}\,M_\odot$/yr and a stellar wind speed of $10^3$ km/s. 
 Thick lines are for the afterglow light-curve calculations including the pair-wind, 
 while thin lines show light-curves obtained if the pair-wind were ignored. Note that 
 the pair-wind brightens the afterglow emission during the first minute by up to
 3 magnitudes, its effect disappearing after the second minute. 
 at the peak.}
\end{figure}

\begin{figure}
\centerline{\psfig{figure=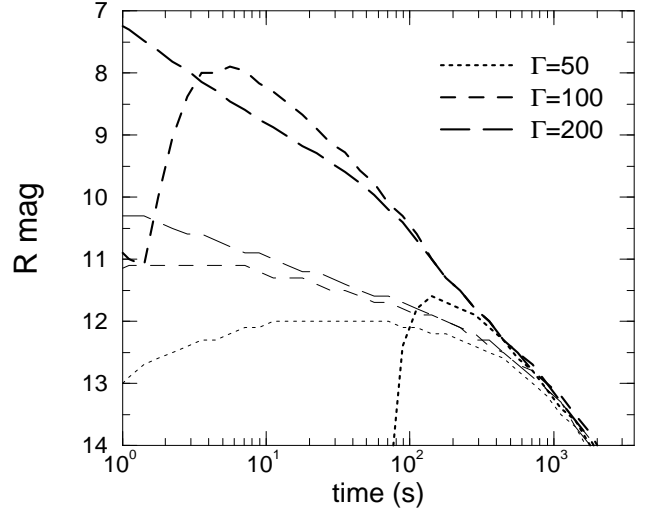,width=8.5cm,height=7cm}}
\caption{ Afterglow light-curves for the parameters given in Figure 6, for
 the ejecta Lorentz factors given in legend and a stellar wind with $A_* = 1$. 
 For $\Gamma = 50$ the pair-wind reaches initially a Lorentz factor $\gamma_\pm$ 
 larger than that of the GRB ejecta and a gap develops between them, the afterglow 
 emission starting when $\gamma_\pm$ has fallen below $\Gamma$ and the ejecta sweep-up
 the pair-wind. As for Figure 6, the effect of the pair-wind on the afterglow emission
 subsides after a couple of minutes.}
\end{figure}

\section{Conclusions}

 Assuming that GRBs are produced by internal shocks at $10^{13}-10^{14}$ cm, a small 
fraction of the gamma-ray photons is converted into electron-positron pairs at a distance 
of less than $r_{pair} = 10^{16}$ cm from the center of explosion. The fractional energy 
carried away by this pair-wind is $\sim 10^{-6} n$, where $n$ is the particle density in 
$\cm3$ of the unperturbed, circumburst medium. The number of pairs created per external 
electron is of order a few hundred and the average Lorentz factor of the pairs is a few. 

 For plausible values of the fireball kinetic energy, electron and magnetic field 
parameters in the shocked circumburst medium, and for an external density $n=100\, \cm3$, 
the optical afterglow from the forward shock peaks around 10 minutes from the burst 
(when the peak of the synchrotron spectrum reaches the optical domain) and at magnitude 
$R \sim 14$ (Figure 3). For a tenfold increase in the density, the peak brightens by one 
magnitude (\eq [\ref{Rp}]) and occurs earlier due to the electron cooling.

 By softening the afterglow spectrum and by increasing the number of radiating particles, 
the pair-wind brightens the rise of the optical light-curve. For densities higher than 
$n \sim 10^3\, \cm3$ the peak optical emission occurs at radii where the pair production 
is significant, leading to a brightening of the optical peak emission (Figure 3) if the 
ejecta Lorentz factor is less than $\siml 300$ (Figure 4). A higher Lorentz factor
suppresses the pair-wind acceleration/enrichment and its brightening of the afterglow 
emission until $t_{pair} = (1+z) r_{pair}/(c \Gamma^2) \sim 10\, (\Gamma/300)^{-2}$ s, but 
leaves unaltered the afterglow emission at later times, past the deceleration timescale.
 
 For conservative values of the afterglow kinetic energy and electron/magnetic field
parameters, the upper limits set by ROTSE and LOTIS on the prompt optical emission 
from several GRBs at $10^2-10^3$ s, ranging from 13 to 16 mag, require that the density 
of the circumburst medium within 0.01 pc of the GRB progenitor is less than $100\,\cm3$. 
This limit is consistent with the typical densities of $0.1-30 \,\cm3$ inferred at 
$r > 10^{17}$ cm from fits to the broadband emission of several GRB afterglows 
(\eg Panaitescu \& Kumar 2002). Moreover, our results show that a wind-like medium 
around the GRB progenitor, corresponding to a mass-loss to wind speed ratio above 
$10^{-5}\, M_\odot/{\rm yr}/ (10^3\, {\rm km/s})$ yields optical counterparts brighter 
than $R = 13$ up to almost $10^3$ s. Thus, the wind of a Wolf-Rayet star is too dense and incompatible with the ROTSE and LOTIS upper limits.

 Figures 3, 4, 6, and 7 show that the brightening of the pair-wind of the early afterglow 
emission is much more prominent for a wind-like external medium. This feature is present 
within the first 100 seconds of the optical afterglow and provides a test for the type of 
circumburst environment, using future Swift observations.

 For $n \simg 10^6\, \cm3$ all photons with energy above the threshold for $e^\pm$ 
production are transformed into pairs and the GRB spectrum is severely modified. 
In this case the pair-wind becomes optically thick to Thomson scattering and the burst 
may not be detectable. However, an optical counterpart resulting from the interaction 
of the ejecta with the pair-wind would be very bright ($R \sim 10$), unless it is 
attenuated by the dust within the dense medium. Such flashes, if produced, would be 
missed by ROTSE and LOTIS, in the absence of the GRB emission.

\section*{Acknowledgments}
We acknowledge many interesting discussions with Bohdan Paczy\'nski, who 
suggested that we investigate the effect of pair-wind on GRB optical afterglows.
We are indebted to the referee, Andrei Beloborodov, for a number of helpful
suggestions that have improved the paper significantly.

\label{lastpage}

\end{document}